\documentstyle[12pt,preprint]{aastex}

\begin{document}
\input psfig
                          
\title{The Evolution of Neutrino Astronomy} 

\author{John N. Bahcall}
\affil{Institute for Advanced Study, Olden Lane, Princeton, NJ 08540}
 \and 
\author{Raymond Davis, Jr.}
\affil{Department of Physics and Astronomy, University of Pennsylvania,
Philadelphia, PA 19104}

How did neutrino astronomy evolve? Are there any useful lessons for
astronomers and physicists embarking on new observational ventures
today? We will answer the first question from our perspective. You, 
the reader, can decide
for yourself whether there are any useful lessons.

\begin{figure}[tb]
\centerline{\psfig{figure=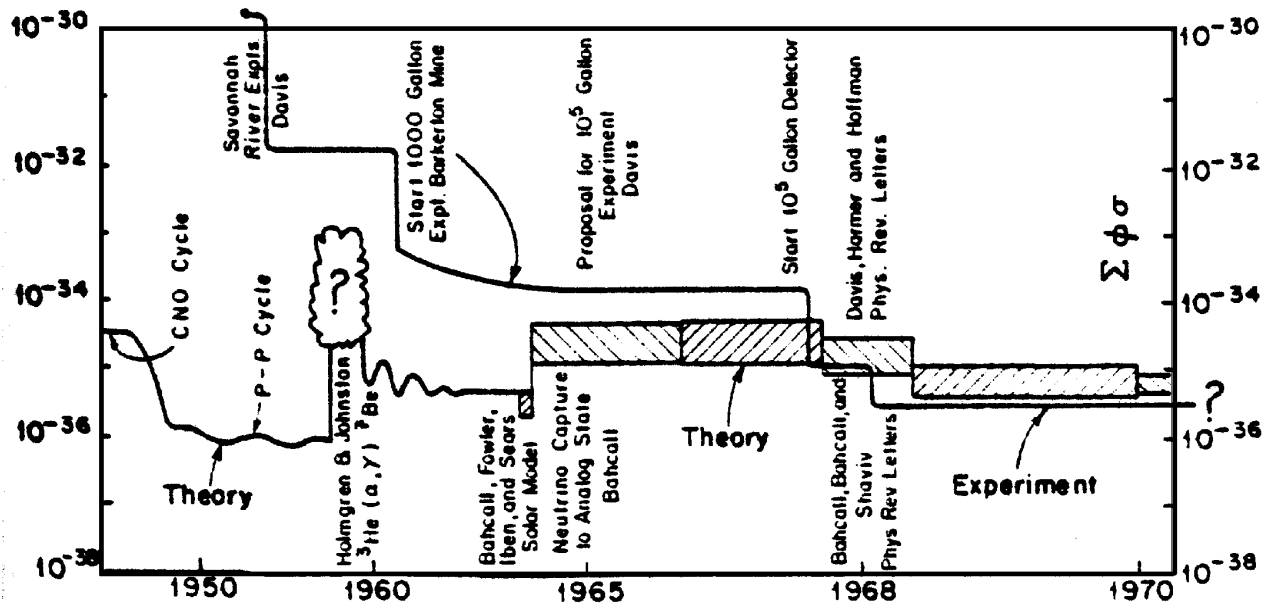,width=6.5in}}
\tightenlines
\caption[]{\small Some of the principal events in the early development of
the solar neutrino problem. The experimental upper limit is indicated
by the think black curve and the range of theoretical values by the
cross-hatched region. The units are captures per target atom per
second ($10^{-36}$ capture/target atom/s $= $ 1 SNU). (Viewgraph:
R. Davis jr., circa 1971.)
\label{fig:davisdrawing}}
\end{figure}

The possibility of observing solar neutrinos began to be discussed
seriously following the 1958 experimental discovery by Holmgren and
Johnston that the cross section for production of the isotope $^7$Be
by the fusion reaction $^3$He + $^4$He $~ \rightarrow ~ ^7{\rm Be} +
\gamma$ was more than a thousand times larger than was previously
believed. This result led Willy Fowler and Al Cameron to suggest that
$^8$B might be produced in the sun in sufficient quantities by the
reaction $^7$Be + $p$ $~ \rightarrow ~ ^8{\rm B} + \gamma$ to produce
an observable flux of high-energy neutrinos from $^8$B beta-decay.
Figure~\ref{fig:davisdrawing} shows the early evolution of neutrino
astronomy as described in a viewgraph from a colloquium given by Ray
at Brookhaven National Laboratory in 1971.

We begin our story in 1964,
when we published back-to-back papers in Physical Review Letters
arguing that a $100,000$ gallon  detector of perchloroethylene
could be built which would measure
the solar neutrino capture rate on chlorine~\footnote{We were asked to
write a brief {\it Millennium Essay for the PASP} on 
 the evolution of neutrino astronomy from our
personal perspective. We present the way the history looks to us 
more than thirty-five years after our collaboration began 
and emphasize those aspects of the
development of neutrino astronomy 
that may be of interest or of use to physicists and 
astrophysicists
today. We stress that all history is incomplete and distorted by
the passage of time and the fading of memories. For earlier more
detailed reviews, the reader can consult two articles we wrote
when
the subject was still in its childhood and  our memories were more
immediate (Bahcall \& Davis 1976, 1982).  The interested reader can find
references in these articles 
to the early works of Bethe, of Holmgren and Johnston, and
of Fowler and Cameron and to the works of  many other 
early pioneers in stellar
fusion and stellar astrophysics.}.  Our motivation was to
use neutrinos to look into the interior of the sun and thereby 
test directly the theory of stellar evolution and nuclear energy generation in
stars. The particular development that made us realize  that the
experiment could be done was the demonstration (by John 
in late 1963) that the
principal neutrino absorption cross section on chlorine was twenty
times larger than previously calculated  due to a super-allowed nuclear
transition to an excited state of argon.

If you have a good idea today, it likely will require many committees, many 
years,  and many people in order to get the project from 
concept to observation. The situation was
very different in 1964. Once the decision to go ahead was made, a very
small team designed and built the experiment; the entire team
consisted of  Ray, Don Harmer
(on leave from Georgia Tech), and John Galvin (a technician who worked
part-time on the experiment). Kenneth Hoffman, a (then) young engineer
provided expert advice on technical questions.
The money came out of the chemistry budget at Brookhaven National
Laboratory. Neither of us
remember a formal proposal ever being written to a funding agency. 
The total capital
expenditure to excavate the cavity in the Homestake Gold Mine in South
Dakota, to build the tank,  and to purchase the
liquid was $0.6$ million dollars (in ~ 1965 dollars).

During the period 1964-1967, Fred Reines and his group worked on three
solar neutrino experiments in which  recoil electrons produced by
neutrino interactions would be detected by observing the associated
light in an organic scintillator. Two of the experiments, which 
exploited the elastic scattering of
neutrinos by electrons, were actually performed and led to a
(higher-than-predicted) upper
limit on the $^8$B solar neutrino flux.
The third experiment, 
which was planned to detect neutrinos absorbed by $^7$Li, was
abandoned after the initial chlorine results showed that the solar
neutrino flux was low.  These three experiments introduced the
technology of organic scintillators into the arena of solar neutrino
research, a technique that will only finally be used in 2001 when the
BOREXINO detector will begin to detect low energy solar
neutrinos. Also during this period, John investigated the properties
of neutrino electron scattering and showed that the forward peaking
from $^8$B neutrinos is large, a feature that was incorporated two
and half decades later in the Kamiokande (and later SuperKamiokande)
water Cherenkov detectors.

\centerline{
\vbox{\hsize=4.75in
\psfig{figure=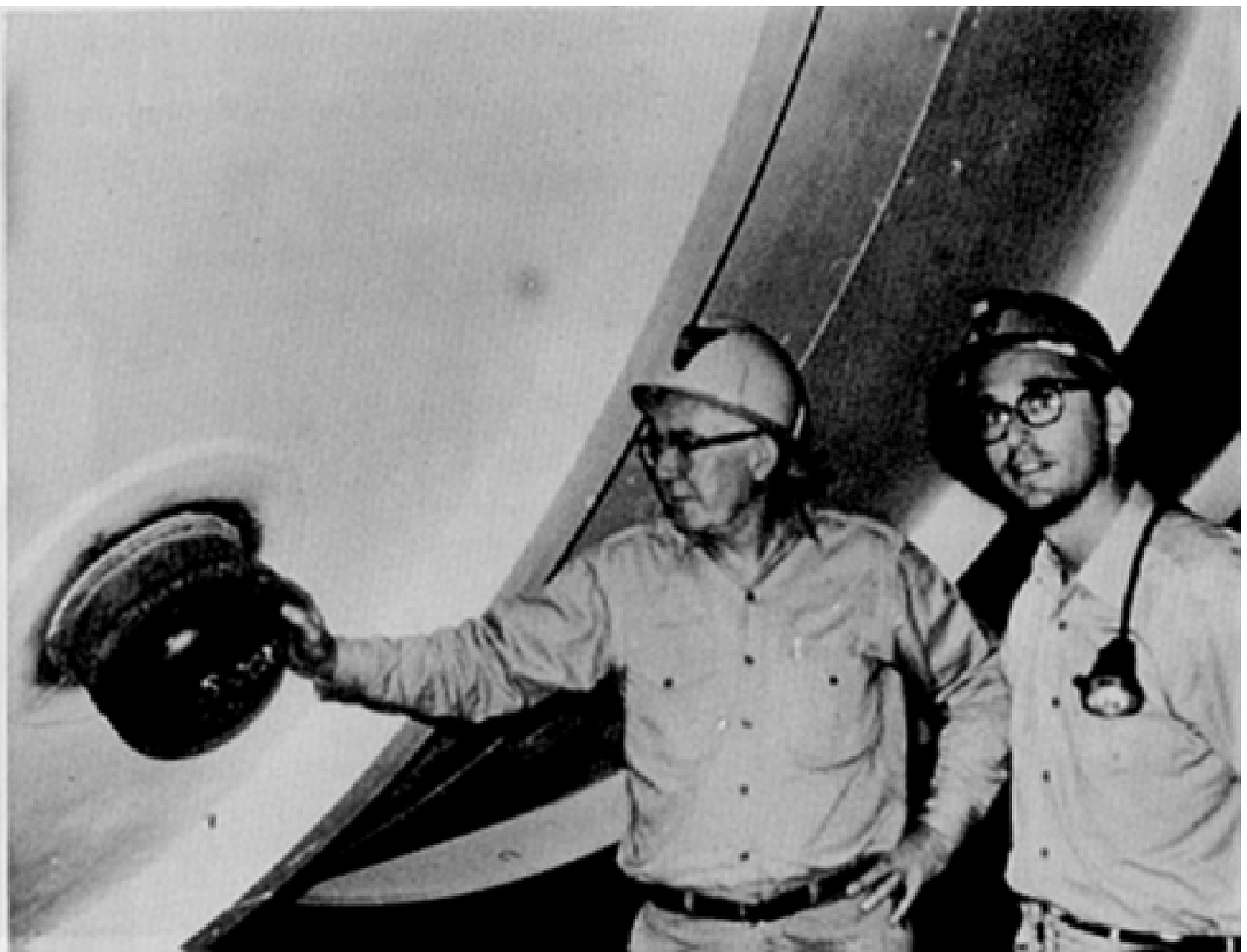,width=4.75in}
\noindent\small Ray Davis shows John Bahcall the tank containing 100,000 gallons of
perchloroethylene.  The picture was taken in the Homestake mine
shortly before the experiment began operating.
}}
\bigskip

The first results from the chlorine experiment were published in 1968,
again in a back-to-back comparison (in PRL) between measurements and
standard predictions. The initial results have been remarkably
robust; the conflict between chlorine measurements and standard solar
model predictions has lasted over three decades.  
The main improvement has been in the slow 
reduction of the
uncertainties in both the experiment and the theory.  
The efficiency of the Homestake chlorine experiment was tested by
recovering carrier solutions, by producing $^{37}$Ar in the tank with
neutron sources, and by recovering $^{36}$Cl inserted in a tank of
perchloroethylene.  The solar model was verified by comparison with
precise helioseismological measurements.

For more than two decades, 
the best-estimates for the observational and for the theoretical
prediction  have remained essentially constant.  The discrepancy
between the standard solar model prediction and the chlorine
observation became widely known as ``the solar neutrino problem.''

Very few people worked on solar neutrinos during the period
1968-1988. The chlorine experiment was the only solar neutrino
experiment to provide data in these two decades.  It is not easy for
us to explain why this was the case; we certainly tried hard to
interest others in doing different experiments and we gave many joint
presentations about what came to be known as ``the solar neutrino
problem''.  Each of us had one principal collaborator during this long
period, Bruce Cleveland (experimental) and Roger Ulrich (solar
models).  A large effort to develop a chlorine experiment in the
Soviet Union was led by George Zatsepin, but it was delayed by the
practical difficulties of creating a suitable underground site for the
detector.  Eventually, the effort was converted into a successful
gallium detector, SAGE, led by Vladimir Gavrin and Tom Bowles, that
gave its first results in 1990.

Only one year after the first (1968) chlorine results were published,
Vladimir Gribov and Bruno Pontecorvo proposed that the explanation of
the solar neutrino problem was that neutrinos oscillated between the
state in which they were created and a more difficult to detect state.
This explanation, which is the consensus view today, was widely
disbelieved by nearly all of the particle physicists we talked to in
those days.  In the form in which solar neutrino oscillations were
originally proposed by Gribov and Pontecorvo, the process required
that the mixing angles between neutrino states be much larger than the
quark mixing angles, something which most theoretical physicists
believed, at that time, was unlikely.  Ironically, a flood of particle
theory papers explained, more or less `naturally', the large neutrino
mixing angle that was decisively demonstrated thirty years later in
the SuperKamiokande atmospheric neutrino experiment.

One of the most crucial events for early solar neutrino research
occurred in  1968 
while we were relaxing in the sun  after a swim at the CalTech
pool.
Gordon Garmire (now a PI for the Chandra X-ray satellite) 
came up to Ray, introduced himself,
and said he had heard about the chlorine experiment.
He suggested to Ray that it might be possible to reduce significantly
the background by using pulse rise time discrimination, a technique
used for proportional counters in space experiments.
The desired fast-rising pulses from $^{37}$Ar Auger electrons are
different from the slower  rising pulses from a background gamma or
cosmic ray.
Ray went back to Brookhaven and asked the local electronic
experts if it would be possible to implement this technique for the
very small counters he used. The initial answer was that the available
amplifiers were not fast enough to be used for this purpose with the
small solar neutrino counters. But,
in about a year  three first class electronic
engineers at BNL, Veljko Radeca,  Bob Chase, and Lee Rogers were able to build
electronics fast enough to be used to measure the rise time in Ray's
counters. 

This `swimming-pool' improvement was crucial for the success
of the chlorine experiment and the subsequent radio-chemical
gallium solar neutrino 
experiments,  SAGE, GALLEX, and GNO. Measurements of the
rise-time as well as the pulse energy greatly reduce the background
for radio-chemical experiments. The backgrounds can be as low as one
event in three months.  

In 1978,
after a decade of disagreement between the Homestake neutrino
experiment and standard solar model predictions, it was clear to
everyone that the subject had reached an impasse and a new experiment
was required. The chlorine experiment is, according to standard solar
model predictions, sensitive primarily to neutrinos from a rare fusion
reaction that involves $^8$B neutrinos. These neutrinos are produced in only
2 of  every $10^4$ terminations of the basic $pp$ fusion
chain. 
In the early part of 1978,
there was a conference of interested scientists who got
together at Brookhaven to discuss what to do next. 
The consensus decision was that we needed an
experiment that was sensitive to the low energy neutrinos from the
fundamental $pp$ reaction. 

The only remotely-practical possibility
appeared to be another radiochemical experiment, this time with
$^{71}$Ga (instead of $^{37}$Cl) as the target. 
But, a gallium experiment (originaly proposed by the Russian theorist
V. A. Kuzmin in 1965) was expensive; we needed about three times
the world's annual production of gallium to do a useful
experiment. 
In an effort to generate enthusiasm for a gallium experiment, 
we wrote another Physical Review Letters paper,  this time
with a number of interested
experimental colleagues. 
We argued that a gallium detector was feasible and that
a gallium measurement, which would be sensitive to the fundamental
$p-p$ neutrinos, would distinguish between broad classes of
explanations for the discrepancy between prediction and observation in
the $^{37}$Cl experiment.
Over the next five or six years, the idea was
reviewed a number of times in the United States, always very
favorably. DOE appointed a blue ribbon panel headed by Glen Seaborg
that endorsed enthusiastically 
both the experimental proposal and the
theoretical justification.

To our great frustration and disappointment, the gallium experiment
was never funded in the United States, although the experimental ideas
that gave rise to the Russian experiment (SAGE) and the
German-French-Italian-Israeli-US experiment (GALLEX) largely
originated at Brookhaven.  Physicists strongly supported the
experiment and said the money should come out of an astronomy budget;
astronomers said it was great physics and should be supported by the
physicists. DOE could not get the nuclear physics and the particle
physics sections to agree on who had the financial responsibility for
the experiment. In a desperate effort to break the deadlock, John was
even the PI of a largely Brookhaven proposal to the NSF (which did not
support proposals from DOE laboratories).A pilot experiment was
performed with 1.3 tons of gallium by an international collaboration
(Brookhaven, University of Pennsylvania, MPI, Heidelberg, IAS,
Princeton, and the Weizmann Institute) which developed the extraction
scheme and the counters eventually used in the GALLEX full scale
experiment.

In strong contrast to what happened in the United States, Moissey
Markov, the Head of the Nuclear Physics Division of the Russian
Academy of Sciences, helped establish a neutrino laboratory within the
Institute for Nuclear Research, participated in the founding of the
Baksan neutrino observatory , and was instrumental in securing $60$
tons of gallium free to Russian scientists for the duration of a solar
neutirno experiment.

The Russian-American gallium experiment (SAGE) went ahead under the
leadership of Vladimir Gavrin, George Zatsepin (Institute for Nuclear
Research, Russia), and Tom Bowles (Los Alamos) and the mostly European
experiment (GALLEX) was led by Till Kirsten (Max Planck Institute,
Germany). Both experiments had a
strong but not primary US participation.

The two gallium experiments were performed in the decade of the 1990's
and gave very similar results, providing the first experimental
indication of the presence of $p-p$ neutrinos.  Both experiments were
tested by measuring the neutrino rate from an intense laboratory
radioactive source.

There were two dramatic developments in the solar neutrino saga, one
theoretical and one experimental, before the gallium experiments
produced observational results.
In 1985, two Russian  physicists proposed an imaginative
solution of the solar neutrino problem that built upon the earlier
work of Gribov and Pontecorvo and, more directly, the insightful
investigation by Lincoln Wolfenstein
(of Carnegie Mellon). Stanislav Mikheyev and Alexei 
Smirnov showed that, if neutrinos
have masses in a relatively wide range, then a resonance phenomenon
in matter
(now universally known as the MSW effect) could convert efficiently 
many of the
electron-type neutrinos created in the interior of the sun to more
difficult to detect muon and tau neutrinos.  
The MSW effect can work for small or large neutrino mixing angles.
Because of the elegance
of the theory and the possibility of explaining the experimental 
results with small
mixing angles (analogous to what happens in the quark sector),
physicists immediately began to be more sympathetic to particle
physics solutions to the solar neutrino problem. More importantly, they
became enthusiasts for new solar neutrino experiments.

The next big break-through also came from an unanticipated
direction. The Kamiokande water Cherenkov detector was developed to
study proton decay in a mine in the Japanese Alps; it set an important
lower limit on the proton lifetime.  In the late 1980's, the detector
was converted by its Japanese founders, Masatoshi Koshiba and Yoji
Totsuka, together with some American colleagues (Gene Beier and Al
Mann of the U. of Pennsylvania) to be sensitive to the lower energy
events expected from solar neutrinos. With incredible foresight, these
experimentalists completed in late 1986 their revisions to make the
detector sensitive to solar neutrinos, just in time to observe the
neutrinos from Supernova 1987a emitted in the LMC 170,000 years
earlier.  (Supernova and solar neutrinos have similar energies, $\sim
10$ MeV, much less than the energies that are relevant for proton
decay.)  In 1996, a much larger water Cherenkov detector (with 50,000
tons of pure water) began operating in Japan under the leadership of
Yoji Totsuka, Kenzo Nakamura, Yoichiro Suzuki (from Japan) , and Jim
Stone and Hank Sobel (from the United States).

So far, five experiments have detected solar neutrinos in
approximately the numbers (within a factor of two or three) and in the
energy range ($ < 15$ MeV) predicted by the standard solar model. This
is a remarkable achievement for solar theory since the $^8$B
neutrinos that are observed primarily in three of these experiments
(chlorine, Kamiokande, and its successor SuperKamiokande) depend upon
approximately the $25$th power of the central temperature.  The same
set of nuclear fusion reactions that are hypothesized to produce the
solar luminosity also give rise to solar neutrinos. Therefore, these
experiments establish empirically that the sun shines by nuclear
fusion reactions among light elements in essentially the way described
by solar models.

Nevertheless, all of the experiments disagree quantitatively  with the
combined predictions of the standard solar model and the standard
theory of electroweak interactions (which implies that nothing much
happens to the neutrinos after they are created). The disagreements
are such that they appear to require some new physics that changes the
energy spectrum of the neutrinos from different fusion sources.

Solar neutrino research today is very different from what it was three
decades ago. The primary goal now is to understand the neutrino
physics, which is a prerequisite for making more accurate tests of the
neutrino predictions of solar models. Solar neutrino experiments today
are all large international collaborations, each typically involving 
of order $10^2$ physicists. Nearly all of the new experiments are
electronic, not radiochemical, and the latest generation of
experiments measure
typically several thousand events per year (with reasonable energy
resolution), compared to rates that were typically 25 to 50 per year
for the radiochemical experiments (which have no energy resolution, only an
energy threshold). Solar neutrino experiments are
currently being carried out in Japan (SuperKamiokande, in the
Japanese Alps), in Canada (SNO, which uses a kiloton of heavy water in 
Sudbury, Ontario), in Italy (BOREXINO, ICARUS, and GNO,  each 
sensitive to a different energy range and all  
operating in the Gran Sasso Underground Laboratory ), in  Russia
(SAGE, in the Caucasus region ), and in the United States (Homestake
chlorine experiment). 
The SAGE, chlorine, and GNO experiments are radiochemical;
the others are electronic.

Since 1985, the chlorine experiment has been operated by the
University of Pennsylvania under the joint leadership of Ken Lande and
Ray Davis.  Lande and Paul Wildenhain have introduced major
improvements in the  extraction and measurement systems,
making the chlorine experiment a valuable source of new precision data.

The most challenging and important frontier for solar neutrino
research is to develop experiments that can measure the energies
of individual low-energy neutrinos from the basic $pp$ reaction, which
constitutes (we believe) more than $90$\% of the solar neutrino flux.

Solar neutrino research is a community activity. 
Hundreds of experimentalists have collaborated to carry  out 
difficult, beautiful  measurements of the elusive neutrinos.
Hundreds of other 
researchers helped refine 
 the solar
model predictions, measuring accurate nuclear and solar 
parameters and calculating input data such as opacities and  equation of
state. 

Three people played special roles. Hans Bethe was
the architect of the theory of 
nuclear fusion reactions in stars, as well as  our mentor and
hero. Willy Fowler was a powerful and enthusiastic supporter of each
new step and his keen physical insight motivated much of what was
done in solar neutrino research. 
Bruno Pontecorvo opened everyone's eyes with his 
original insights, including his early discussion of the 
advantages of using chlorine as a neutrino detector and his
suggestion that neutrino oscillations might be important.

In the next decade, neutrino astronomy will move beyond our cosmic
neighborhood and, we hope, will detect distant sources.  The most likely
candidates now appear to be gamma-ray bursts.  If the standard fireball
picture is correct and if gamma-ray bursts produce the observed
highest-energy cosmic rays, then very high energy ($~ 10^{15}$ eV)
neutrinos should be observable with a ${\rm km^2}$
detector. Experiments with the capability to detect
neutrinos from
gamma-ray bursts are being developed at the South Pole (AMANDA
and ICECUBE), in the Mediterranean Sea (ANTARES, NESTOR) and even in
space. 

Looking back on the beginnings of solar neutrino 
astronomy, one lesson appears clear to us: if you can measure
something new with  reasonable accuracy, then 
you have a chance to discover something
important. The history of astronomy shows that very likely what you
will discover
is not what you were looking for. It helps to be lucky.

\end{document}